\documentclass[12pt,a4paper]{article}
\usepackage{epsfig,epsf,mathrsfs,amssymb,amsmath,amsfonts,psfrag}
\usepackage[hang,small,bf]{caption}

\renewcommand*{\today}{
    \number\day \ifcase\day \or
    st\or nd\or rd\or th\or th\or th\or th\or th\or th\or th\or 
    th\or th\or th\or th\or th\or th\or th\or th\or th\or th\or 
    st\or nd\or rd\or th\or th\or th\or th\or th\or th\or th\or 
    st\fi\space \ifcase\month \or
    January\or February\or March\or April\or May\or June\or 
    July\or August\or September\or October\or November\or
    December\fi \space\number\year}

\def\feynsl#1{
  \setbox0=\hbox{/} \setbox1=\hbox{$#1$}
  \dimen0=\wd0 \advance\dimen0 by -\wd1 \divide\dimen0 by 2
  \ifdim\wd0>\wd1 \raise.15ex\copy0\kern-\wd0\kern\dimen0\copy1\kern\dimen0
  \else \kern-\dimen0\raise.15ex\copy0\kern-\dimen0\kern-\wd1\copy1\fi}

\makeatletter
\def\@eqnnum{\hbox{\reset@font\rm(\theequation)}}
\let\make@eqnnum=\@eqnnum %
\def\eqnum#1{\dec@eqnnum \global\def\make@eqnnum{\reset@font\rm(#1)}%
\def\@currentlabel{#1}%
}
\def\inc@eqnnum{\addtocounter{equation}{1}}
\def\dec@eqnnum{\addtocounter{equation}{-1}}
\@definecounter{equation}%
\@addtoreset{equation}{section} %
\def\theequation@prefix{{\thesection}.} %
\def\theequation{\theequation@prefix\arabic{equation}}%
\makeatother

\begin{document}
\topskip 1cm
\begin{titlepage}
\hspace*{\fill}\parbox[t]{4cm}{
IPPP/03/37\\
DCPT/03/74\\ 23rd July 2003}
\vfill
\begin{center}
{\Large\bf Hadronic Radiation Patterns in\\[7pt] Vector Boson Fusion Higgs Production}\\
\vspace{1.cm}

{V. A. Khoze, W. J. Stirling and P. H. Williams}\\
\vspace{.2cm}
{Institute for Particle Physics Phenomenology\\
University of Durham\\
Durham, DH1 3LE, U.K.}\\

\vspace{.5cm}

\begin{abstract}
  We consider the hadronic radiation patterns for the generic process of $b
  \bar b + 2\; $forward jet production at the LHC, where the (centrally
  produced) $b \bar b$ originate either from a Higgs, a $Z$ or from standard
  QCD production processes. A numerical technique for evaluating the
  radiation patterns for non-trivial final states is introduced and shown to
  agree with the standard analytic results for more simple processes.
  Significant differences between the radiation patterns for the Higgs signal
  and the background processes are observed and quantified. This suggests
  that hadronic radiation patterns could be used as an additional diagnostic
  tool in Higgs searches in this channel at the LHC.
\end{abstract}
\end{center}
 \vfill
\end{titlepage}
\section{Introduction}

The distribution of soft hadrons or jets accompanying energetic final--state
particles in hard scattering processes is governed by the underlying colour
dynamics at short distances \cite{DKT,DKMT,book,emw}. The soft hadrons paint
the colour portrait of the parton hard scattering, and can therefore act as a
`partonometer'
\cite{DKT,DKMT,book,emw,DKTSJNP,DKS,MarWeb,ZEPPEN,Ell96,Kho97,Amu98,klo}.
Since signal and background processes at hadron colliders can have very
different colour structures (compare for example the $s$--channel colour
singlet process $g g \to H \to b \bar b$ with the colour octet process $q
\bar q \to g^* \to b \bar b$), the distribution of accompanying soft hadronic
radiation in the events can provide a useful additional diagnostic tool for
identifying new physics processes.\par Examples that have been studied in the
literature in this way include Higgs \cite{Heyssler:1998wm}, $Z'$
\cite{Ell96} and leptoquark \cite{Hey97} production.  In each case the new
particle production process was shown to have its own particular colour
footprint, distinctively different from the corresponding background process.
\par Quite remarkably, because of the property of Local Parton Hadron Duality
(see for example Refs.~\cite{DKMT,book,adkt}) the distribution of soft
hadrons can be well described by the amplitudes for producing a single
additional soft gluon.  The angular distribution of soft particles typically
takes the form of an `antenna pattern' multiplying the leading--order hard
scattering matrix element squared. Confirmation of the validity of this
approach comes from studies of the production of soft hadrons and jets
accompanying large $E_T$ jet and $W+$jet production by the CDF \cite{CDF} and
D0 collaborations \cite{D0} at the Fermilab Tevatron.\par One of the most
important physics goals of the CERN LHC $pp$ collider is the discovery of the
Higgs boson. Many scenarios, corresponding to different production and decay
channels, have been investigated, see for example the studies reported in
Refs.~\cite{Carena:2002es,ATLAS,CMS}.  In a recent paper~\cite{Khoze:2002fa},
we have studied Higgs production via vector boson fusion at the LHC, $qq\to
qqH$, where the colour--singlet nature of the $V^*V^*\to H$ production
process naturally gives rise to rapidity gaps between the centrally produced
Higgs and the forward jets\footnote{Another topical example concerns central
  production of new heavy objects (Higgs, SUSY particles etc.) at hadron
  colliders in events with double rapidity gaps,
  $p\overset{(-)}{p}\rightarrow X + M + Y$ (where $+$ indicates a rapidity
  gap), which are caused by the pomeron exchanges in the $t$-channel. For a
  recent discussion and a list of references see~\cite{KMR}.}.  The most
delicate issue in calculating the cross section for processes with these
rapidity gaps concerns the soft survival factor $\hat{S}^2$. This
non-universal factor has been calculated in a number of models for various
rapidity gap processes, see for example~\cite{KKMR} and references therein.
Although there is reasonable agreement between these model expectations, it
is always difficult to guarantee the precision of predictions which rely on
soft physics.  However, in~\cite{Khoze:2002fa} we argued that the
calculations of $\hat{S}^2$ can be checked experimentally by computing and
measuring the event rate for a suitable calibrating process, for example the
production of a $Z$ boson with the same rapidity gap and jet configuration as
for the (comparatively light) Higgs, see also~\cite{Chehime:1992ub}.\par In
this paper we adopt a different approach to the same problem. Rather than
considering the case where the emission of soft hadrons between the jets and
central Higgs is suppressed (rapidity gaps), we instead discuss the inclusive
distribution using the antenna pattern approach to quantify the relative
amounts of soft hadron emission in the signal and background events. In other
words, we quantify how `quiet' the signal events are compared to the
otherwise irreducible background events.\par Thus we have in mind the
following type of scenario.  Suppose an invariant mass peak is observed in a
sample of (tagged) $b \bar b$ events in which there are energetic forward
jets, typical of the vector--boson production process. If such events do
indeed correspond to Higgs production, then the distribution of accompanying
soft radiation in the event --- which we take to mean the angular
distribution of hadrons or `minijets' with energies of at most a few GeV,
well separated from the beam and final--state energetic jet directions ---
will look very different from that expected in background QCD production of
$b \bar b + 2\; $jet events.  Again, the analogous process of $Z (\to b \bar
b) + 2\;$jet production can be used to calibrate the analysis, since these
events are, as we shall see, also generally quieter than the QCD
backgrounds.\par Thus in this study we will consider the hadronic radiation
patterns for the generic process of $b \bar b + 2\; $forward jet production,
where the (central) $b \bar b$ originate either from a Higgs, a $Z$ or from
standard QCD production processes. We will chose configurations (i.e. cuts on
the rapidities and tranvserse momenta of the final--state particles) that
maximise the Higgs signal to background ratio, see~\cite{Khoze:2002fa}.\par
The paper is organised as follows. In the following section we consider the
antenna patterns for Higgs and $Z$ production accompanied by two forward
jets.  We show that for these colour---singlet production processes, fairly
simple analytic expressions can be derived. However this is not the case for
the more complicated QCD background processes. In Section~3 we show that the
radiation patterns for these can be calculated using a more general numerical
technique, which indeed can be applied to arbitrarily complicated processes.
Section~4 summarises our results and presents our conclusions.
\section{Hadronic Antenna Patterns for Higgs and $Z+2$~Jet Production}
\subsection{Higgs and Electroweak $Z$ Production}
\begin{figure}[ht]
  \begin{center}
    \scalebox{1}{\includegraphics{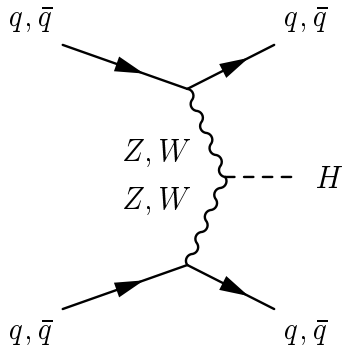}}
    \caption{Higgs production via electroweak vector boson fusion.}
    \label{fig:wwfus}
  \end{center}
\end{figure}

The signal process we are interested in is Higgs production via vector boson
fusion, shown in Fig.~\ref{fig:wwfus}, with subsequent decay of the Higgs 
to $b\bar{b}$.
Furthermore, we restrict our considerations to the case where the outgoing
quark jets are forward in rapidity and the Higgs decay products are central
in the detector. Throughout this paper, we work in the zero width
approximation for the Higgs and $Z$. As vector boson fusion involves no
colour flow in the $t$-channel, the radiation pattern is simply that of the
$2\rightarrow 2$ process $qq^{\prime}\rightarrow qq^{\prime}$, with an
additional colour disconnected $b\bar{b}$. These were calculated
in~\cite{Ell96}. Note also that we work with massless quarks. The radiation
pattern is defined as the ratio of the $2\rightarrow n+1$ and $2\rightarrow
n$ matrix elements using the soft gluon approximation in the former. The
dependence on the soft gluon momentum $k$ then enters via the eikonal factors
(`antennae')~\cite{DKT,book}
\begin{equation}
  [ij] \equiv \frac{p_{i}\cdot p_{j}}{(p_{i}\cdot k)(p_{j}\cdot k)}.  
\end{equation}
For the signal $q(p_1)\,q^{\prime}(p_2)\rightarrow q(p_3)\,q^{\prime}(p_4)+g(k)$
we have
\begin{equation}
  \overline{\sum}|\mathcal{M}|^{2} = \frac{g_{s}^{6}C_{F}}{N_{c}}\left(\frac
           {s^{2}+u^{2}}{t^{2}}\right)2C_{F}([13]+[24])
\end{equation}
with $s \equiv (p_1+p_2)^2$, $t\equiv(p_1-p_3)^2$, $u\equiv(p_1-p_4)^2$.

\begin{figure}[htbp]
  \centering
  \epsfig{file=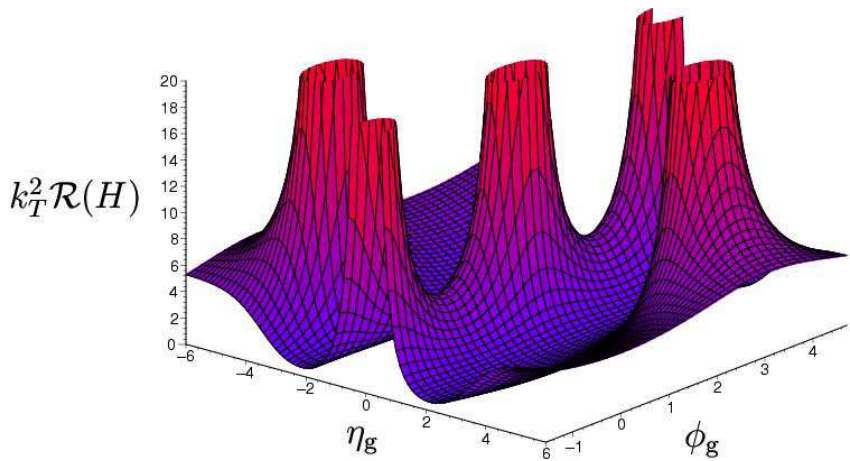,height=60mm}
  \caption{Antenna pattern for $qq^{\prime}\rightarrow qq^{\prime}H\,;\,H\rightarrow b\bar{b}$. Here $\eta_{\textrm{jet}}=3.5$.}
  \label{fig:hant_plot}    
\end{figure}

We then normalise this by the matrix element for the leading order process
$q(p_1)\,q^{\prime}(p_2)\rightarrow q(p_3)\,q^{\prime}(p_4)$:
\begin{equation}
  \overline{\sum}|\mathcal{M}|^{2} = \frac{g_{s}^{4}C_{F}}{N_{c}}
  \left(\frac{s^{2}+u^{2}}{t^{2}}\right).
\end{equation}
Note that in this particular case, the $2 \rightarrow 3$ matrix element in the
soft gluon limit factorises into the form $(2\rightarrow n \textrm{ matrix
  element})\times(\textrm{antenna factor})$. This feature is {\it not}
universal, being restricted to only very simple cases such as this.
The antenna pattern is then
\begin{equation}
  \mathcal{R}(qq^{\prime}\rightarrow qq^{\prime})=g_{s}^{-2}
  \frac{|\overline{\mathcal{M}}_{3}|^{2}(qq^{\prime}
    \rightarrow qq^{\prime}+g)}{|\overline{\mathcal{M}}_{2}|^{2}(qq^{\prime}
    \rightarrow qq^{\prime})}=2C_{F}([13]+[24]).
\end{equation}
As we are working in the zero width approximation\footnote{Actually our
  analysis is formally correct provided that $\Gamma_H \ll E_g$ where $E_g$
  is the typical soft gluon/hadron energy, i.e. the Higgs lives long enough
  to prevent any interference between gluon emission before and after the
  Higgs decays. In any case, such interference would occur only in higher
  orders in $\alpha_{s}$ and is colour suppressed.} we can include the decay
of the Higgs into (massless) $b\bar{b}$ by simply adding the antenna for this
colour disconnected part.  The hadronic radiation pattern for $q(p_1)\,
q^{\prime}(p_2)\,\rightarrow q(p_3)\,q^{\prime}(p_4)\,H\;;\; H\rightarrow
b(p_5)\,\bar{b}(p_6)$ is then
\begin{equation}\label{eq:hant}
  \mathcal{R}(H)=2C_{F}([13]+[24]+[56]).
\end{equation}
In order to visualise the pattern we must specify the kinematics and pick a 
relevant configuration for the incoming and outgoing particles. We label the
four-momenta by
\begin{equation}
  a(p_1) + b(p_2) \rightarrow c(p_3)+d(p_4) + \cdots + g(k),
\end{equation}
where the gluon is soft relative to the other large-$E_T$ final state
partons, i.e. $k\ll E_T$. We ignore the gluon momentum in the energy-momentum
constraints, work in the overall parton centre of momentum frame, fix the Higgs
to be at rest in that frame and its decay products at
$(\eta_{\textrm{b}},\phi_{\textrm{b}})=(0,\pi/2)$ and $(0,3\pi/2)$. With the
notation $p^{\mu}=(E,p_{x},p_{y},p_{z})$, the momenta are then
\begin{eqnarray}\label{eq:4mom}
  p_{1}^{\mu}&=&(m_{H}/2+E_{T}\cosh\eta_{\textrm{jet}},0,0,m_{H}/2+E_{T}\cosh\eta_{\textrm{jet}})
  ,\nonumber\\
  p_{2}^{\mu}&=&(m_{H}/2+E_{T}\cosh\eta_{\textrm{jet}},0,0,-m_{H}/2-E_{T}\cosh\eta_
  {\textrm{jet}}),\nonumber\\
  p_{3}^{\mu}&=&(E_{T}\cosh\eta_{\textrm{jet}},0,E_{T},E_{T}\sinh\eta_{\textrm{jet}}),\nonumber\\
  p_{4}^{\mu}&=&(E_{T}\cosh\eta_{\textrm{jet}},0,-E_{T},-E_{T}\sinh\eta_{\textrm{jet}})
  ,\nonumber\\
  p_{H}^{\mu}&=&(m_{H},0,0,0),\nonumber\\
  p_{b}^{\mu}&=&(m_{H}/2,m_{H}/2,0,0),\nonumber\\
  p_{\bar{b}}^{\mu}&=&(m_{H}/2,-m_{H}/2,0,0),\nonumber\\
  k^{\mu}&=&(k_{T}\cosh\eta_{\textrm{g}},k_{T}\sin\phi_{\textrm{g}},k_{T}
  \cos\phi_{\textrm{g}},k_{T}\sinh\eta_{\textrm{g}}).
\end{eqnarray}
This is the appropriate form for studying the angular distribution of the
soft gluon, parametrised by $\eta_{\textrm{g}}$ and
$\phi_{\textrm{g}}$. Using the kinematics of Eq.~(\ref{eq:4mom}) with 
Eq.~(\ref{eq:hant}) gives
\begin{eqnarray}
  \label{eq:lab_hant}
  \mathcal{R}(H)&=& \frac{2C_{F}}{k_{T}^{2}}\left\{\frac{\cosh\eta_
      {\textrm{jet}}-\sinh\eta_{\textrm{jet}}}
    {(\cosh\eta_{\textrm{g}}-\sinh\eta_{\textrm{g}})(\cosh\eta_{\textrm{jet}}
      \cosh\eta_{\textrm{g}}
      -\cos\phi_{\textrm{g}}-\sinh\eta_{\textrm{jet}}\sinh\eta_
      {\textrm{g}})}\right.\nonumber\\
  &+&\frac{\cosh\eta_{\textrm{jet}}-\sinh\eta_
    {\textrm{jet}}}{(\cosh\eta_{\textrm{g}}+
    \sinh\eta_{\textrm{g}})(\cosh\eta_{\textrm{jet}}\cosh\eta_{\textrm{g}}
    +\cos\phi_{\textrm{g}}+\sinh\eta_{\textrm{jet}}\sinh\eta_{\textrm{g}})}
  \nonumber\\
  &+&\left.
    \frac{2}{(\cosh\eta_{\textrm{g}}-\sin(\phi_{\textrm{g}}+\pi))
      (\cosh\eta_{\textrm{g}}+\sin(\phi_{\textrm{g}}+\pi))}\right\}
\end{eqnarray}
 Note that the result is
independent of $E_{T}$ and $m_{H}$ and that collinear singularities are
situated at $(\eta_{\textrm{g}},\phi_{\textrm{g}}) = (\eta_{\textrm{jet}},
\pi)\,,(-\eta_{\textrm{jet}},0)\,,(0,\pi/2)\,\textrm{and}\,(0,3\pi/2)$.  As
an illustration, Figure~\ref{fig:hant_plot} shows $k_{T}^{2}{\cal R}(H)$ with
$\eta_{\textrm{jet}}=3.5$.
\begin{figure}[htbp]
  \centering \psfrag{x}[t][bl]{$\eta_{\textrm{jet}}$}
  \psfrag{y}[br][bl]{$k_{T}^{2}{\cal R}(qq^{\prime}\rightarrow qq^{\prime})$}
  \epsfig{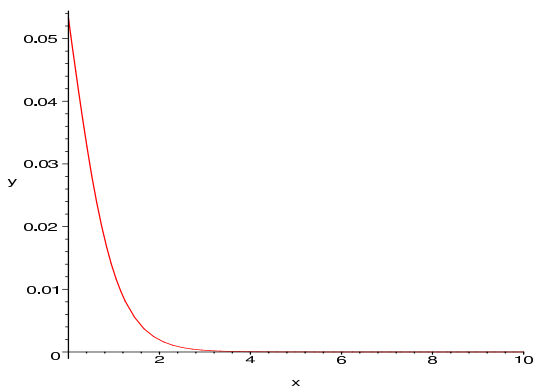}
  \caption{The point $(\eta_{\textrm{g}},\phi_{\textrm{g}})=(0,\pi/2)$ in
    ${\cal R}(qq^\prime\rightarrow qq^\prime)$ as one varies
    $\eta_{\textrm{jet}}$. As the jets move apart, the antenna falls to zero.}
  \label{fig:hant_var_eta}    
\end{figure}
One can clearly see that a colour connection exists between the initial state
parton $p_1$ and final state jet $p_3$, similarly with $p_2$ and $p_4$, and
also between the $b$-quark jets. The antenna pattern is small between the
jets and the $b$'s as there is no colour connection between these --- this is
the `rapidity gap' phenomenon. The emission of soft gluons in the rapidity
gaps decreases as the gap widens. This is illustrated in the case without the
$b$-quark antenna (Fig.~\ref{fig:hant_var_eta}), which shows the antenna
pattern at $(\eta_{\textrm{g}},\phi_{\textrm{g}})=(0,\pi/2)$ as a function of
$\eta_{\textrm{jet}}$.
\par
\begin{figure}[htbp]
  \centering
  \epsfig{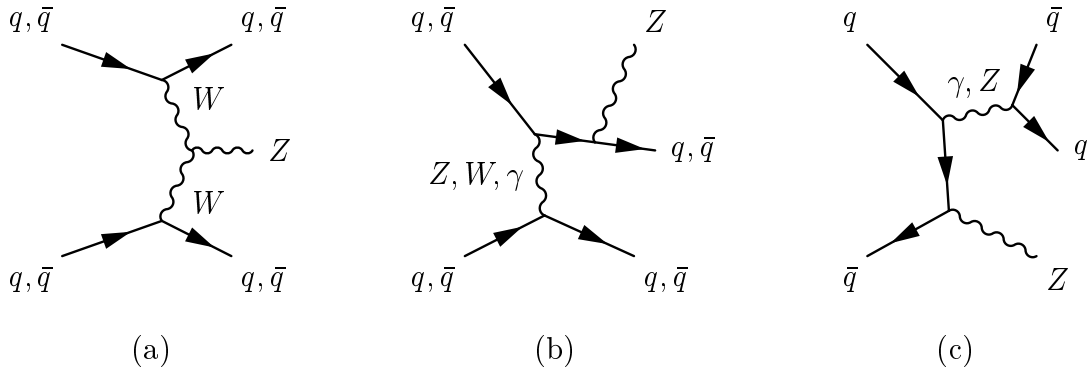}
  \caption{Electroweak $Z$ production.}
  \label{fig:zew}
\end{figure}
Next we consider the analogous electroweak $Z$ production process
(Fig.~\ref{fig:zew}), which can in principle be used to calibrate the Higgs
production process. 
In this case the variety of diagrams at leading order means that there is no
exact eikonal factorisation. However in the kinematic limit we are interested
in --- forward jets and central $Z$ production --- the dominant amplitude is
again the one involving $t$--channel $W$ exchange, i.e. $WW \to Z$, and the
antenna pattern is trivially identical to that for Higgs production.  We will
prove this result, and consider its implications, when we discuss how to
calculate antenna patterns numerically below.
\par
\subsection{QCD $Z$ Production}
\begin{figure}[htbp]
  \centering
  \epsfig{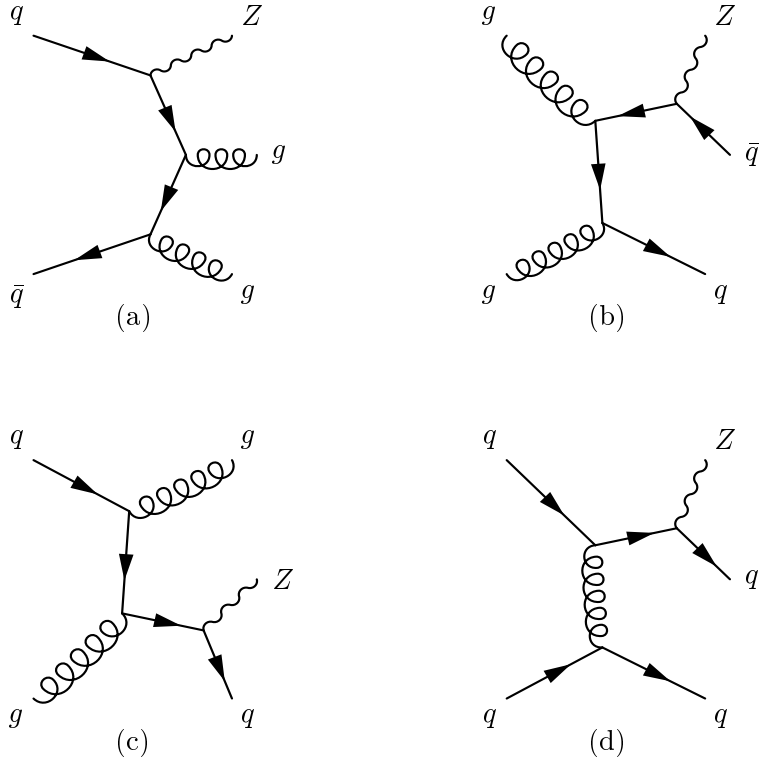}
  \caption{QCD $Z + 2$ forward jet production.}
  \label{fig:zqcd}
\end{figure}
In practice, $Z+ 2\; $jet production can also occur by ${\cal O}(\alpha_S^2
\alpha_W)$ QCD production involving $t$-channel gluon exchange, see
Fig.~\ref{fig:zqcd}d. Because of the different colour structure of such
diagrams we would expect a very different antenna pattern.  

\begin{figure}[htbp]
  \centering
  \epsfig{file=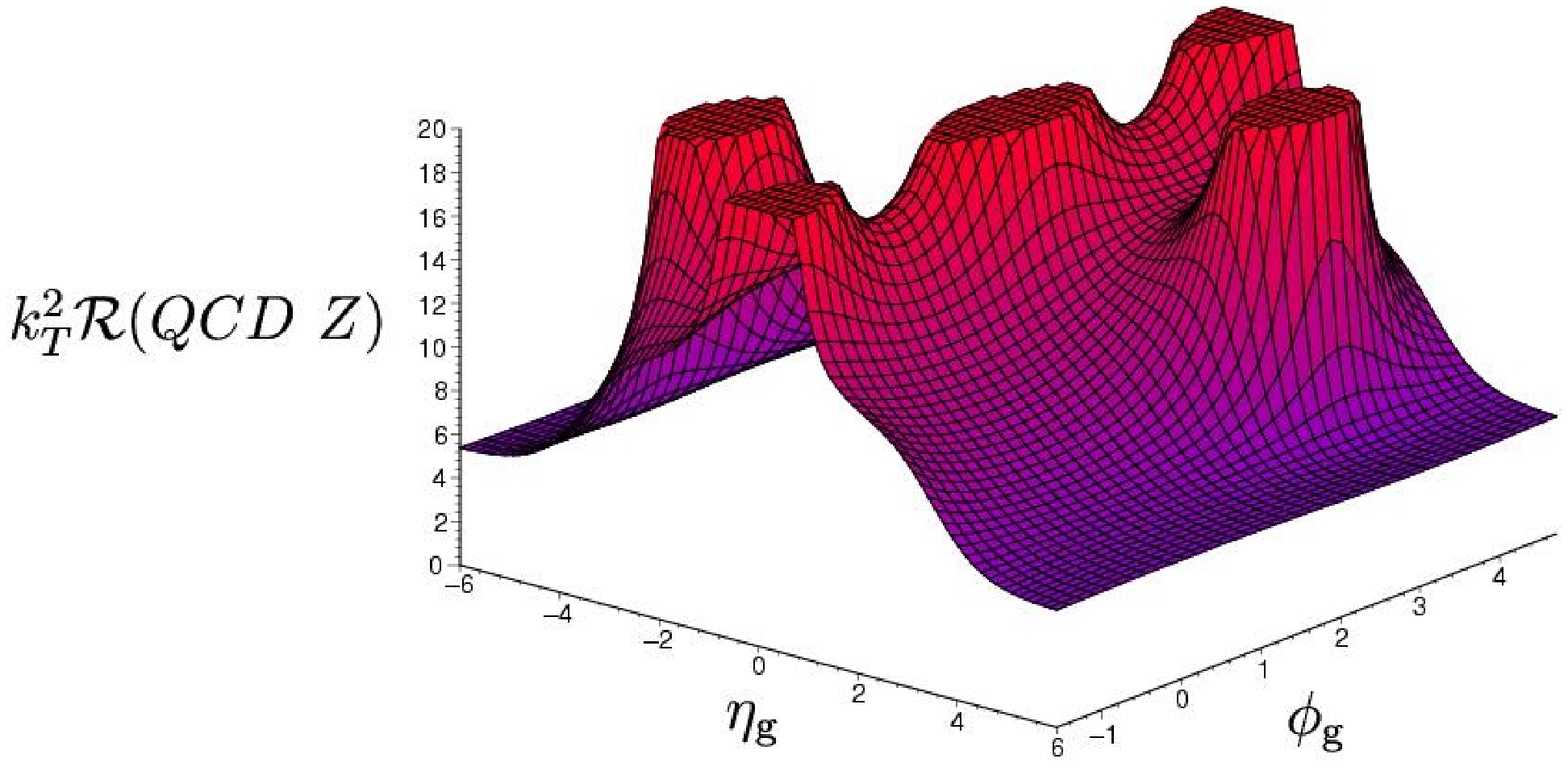,height=60mm}
  \caption{Antenna pattern for $qq^{\prime}\rightarrow qq^{\prime}Z\,\textrm{(QCD)}\,;\,Z\rightarrow b\bar{b}$ with $\eta_{\textrm{jet}}=3.5$.}
  \label{fig:zqcdant}
\end{figure}

\begin{figure}[htbp]
  \centering
  \epsfig{file=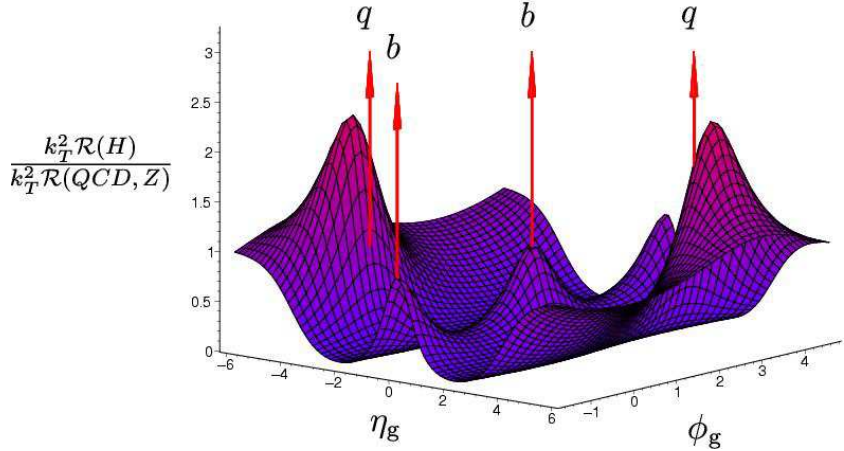,height=60mm}
  \caption{Ratio of Higgs to QCD $Z$ production antenna patterns. The ratio
    is unity at the position of the jets (indicated by arrows) and in the
    forward and backward limits $|\eta_{\textrm{g}}|\rightarrow\infty$.}
  \label{fig:zqcd_h_ratio}
\end{figure}

Once again there is no exact factorisation of an overall soft gluon form
factor and therefore no simple expression for the radiation pattern. However,
as for electroweak $Z$ production the factorisation is restored in the
forward jet -- central $Z$ limit, in which case the antenna pattern is
identical to that for the QCD ${\cal O}(\alpha_S^2)$ $q q' \to q q'$
production process~\cite{Ell96}, i.e.
\begin{equation}
  \label{eq:qcdz}
  \mathcal{R}(QCD\,\,Z) \to 2C_{F}([14]+[23])+\frac{1}{N_{c}}[12;34]+2C_{F}[56],
\end{equation}
where
\begin{equation}
  [ij;kl] \equiv 2[ij]+2[kl]-[ik]-[il]-[jk]-[jl].
\end{equation}
Substituting the kinematics of Eq.~(\ref{eq:4mom}) and plotting the resulting
analytic expression with $\eta_{\textrm{jet}}=3.5$, one obtains
Figure~\ref{fig:zqcdant}.
Before commenting on the differences, we note that both
Figures~\ref{fig:hant_plot} and~\ref{fig:zqcdant} exhibit the same limiting
behaviour
\begin{equation}
  \label{eq:eta_lim}
  \lim_{|\eta_{g}|\rightarrow\infty}k_{T}^{2}\mathcal{R}(H,QCD\,\,Z)=4C_{F},
\end{equation}
as a consequence of both processes having initial state quarks\footnote{Of
  course $qg \to Zqg$ also contributes to $Z+2\;$jet production, and this
  will have a different colour structure from $qq\to Zqq$. For purposes of
  comparison with the Higgs case, we only consider quark induced production
  in this section.}. They are also identical as one approaches the collinear
singularities corresponding to the final state $b$--jets:
\begin{equation}
  \label{eq:jet_lim}
  \lim_{(\eta_{\textrm{g}},\phi_{\textrm{g}})\rightarrow(\eta_{\textrm{jet}},
    \phi_{\textrm{jet}})}k_{T}^{2}\mathcal{R}(H,QCD\,\,Z)\rightarrow 
  4C_{F}\frac{1}{\cosh^{2}
    (\eta_{\textrm{g}}-\eta_{\textrm{jet}})-\cos^{2}(\phi_{\textrm{g}}-
    \phi_{\textrm{jet}})}.
\end{equation}
The difference in the colour flow shows up in the region between the two
final state forward quark jets, as expected. Taking the ratio of the two
patterns makes this difference plain (Fig.~\ref{fig:zqcd_h_ratio}). The
maximum difference occurs at $(\eta_{\textrm{g}},\phi_{\textrm{g}})=(-4.4,0)$
and $(\eta_{\textrm{g}},\phi_{\textrm{g}})=(4.4,\pi)$ when the ratio attains
the value $2.3$. This shows the colour connection between the initial state
(at $\eta_{\textrm{g}}=\pm\infty$) and the forward jets in the Higgs
production case that is suppressed by a factor ${\cal
  O}(\frac{1}{N_{c}^{2}})$ in the QCD Z-production case. Another interesting
phase space point is at $(\eta_{\textrm{g}},\phi_{\textrm{g}})=(0,0)$, i.e.
the central region transverse to the $b \bar b$ axis. Here the radiation
pattern increases by a factor of three going from Higgs to QCD $Z$
production, indicating the presence of an additional underlying colour
connection in the latter case.

\section{Numerical Hadronic Antenna Patterns}
\begin{figure}[htbp]
  \centering
  \epsfig{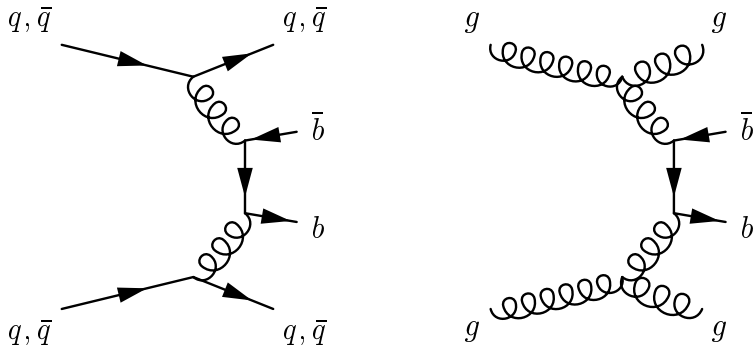}
  \caption{Examples of QCD dijet plus $b\bar{b}$ production diagrams.}
  \label{fig:qcdbbb}
\end{figure}

An important (and dominant) background to the processes considered in the
previous section comes from QCD ${\cal O}(\alpha_S^3)$ $b \bar b + 2\;$ jet
production when $M_{b\bar b} \sim M_{H,Z}$\footnote{We are not discussing
  here the background caused by a possible misidentification of the gluons
  as $b$ jets. For a recent treatment of this see~\cite{DeRoeck:2002hk}.}. 
Some sample diagrams are shown in Fig.~\ref{fig:qcdbbb}.
There is clearly no unique and simple colour flow associated with these
diagrams, and hence no compact analytic antenna pattern can be derived. This
is an example of a situation where there is no factorisation of the form
$(2\rightarrow 4 \textrm{ matrix element})\times(\textrm{antenna factor})$.
However we can instead use a purely numerical method in which we compare the
values of the $2\rightarrow n$ and $2\rightarrow n+1$ matrix elements at each
point in phase space, their ratio in the soft gluon limit defining the
antenna pattern. In order to verify that this methodology works, and in
particular to establish how soft the gluon has to be before the limiting
pattern is reached to some level of precision, we first make a numerical
evaluation of the analytic radiation patterns discussed in the previous
section.
\subsection{Comparison of Numerical and Analytic Antenna Patterns for Signal
  Processes}
\begin{figure}[htbp]
  \centering
  \epsfig{file=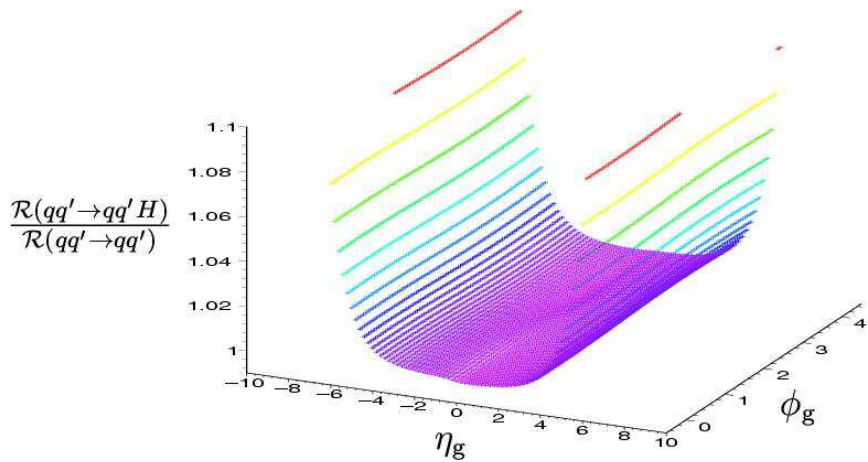,height=60mm}
  \caption{Ratio of numerical $qq^{\prime}\rightarrow qq^{\prime}H$ to 
    analytic $qq^{\prime}\rightarrow qq^{\prime}$ antenna patterns 
    with $|\eta_{\textrm{jet}}|=2$ and $k_{T\textrm{g}}=1$~GeV.}
  \label{fig:qqh_anrat_y2_kt1}
\end{figure}

\begin{figure}[htbp]
  \centering 
  \epsfig{file=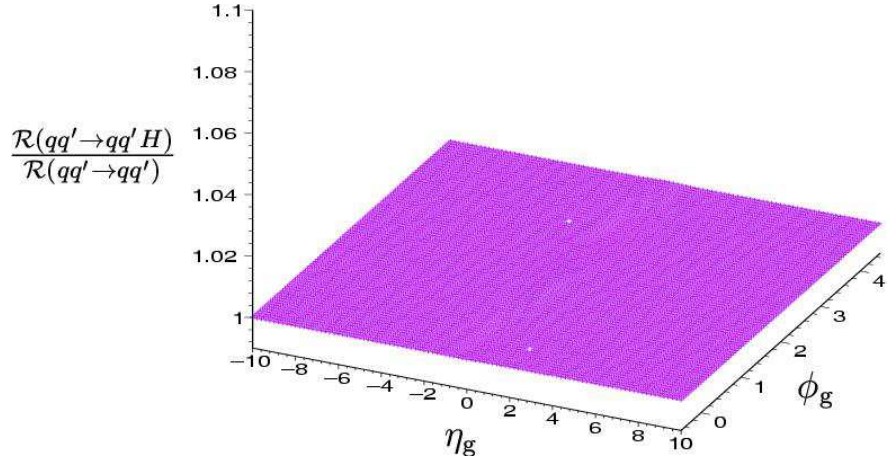,height=60mm}
  \caption{Ratio of numerical $qq^{\prime}\rightarrow qq^{\prime}H$ to analytic $qq^{\prime}\rightarrow qq^{\prime}$ antenna patterns with $|\eta_{\textrm{jet}}|=2$ and $k_{T\textrm{g}}=10^{-5}$~GeV.}
  \label{fig:qqh_anrat_y2}
\end{figure}

Unlike the analytic case, where we can simply ignore the momentum of
the soft gluon in assigning a kinematic configuration that respects momentum
conservation, we must account for the numerically finite gluon momentum in
evaluating the matrix elements. Thus there is a degree of arbitrariness
introduced. We choose to assign the momenta such that the central boson or
$b\bar{b}$ system cancels the $3$-momentum of the soft gluon. In other words
\begin{equation}
  \label{eq:zmom}
  p^{\mu}_{Z,H,b\bar{b}}=\left(\sqrt{m_{Z,H}^2+k^{2}},-\underline{k}\right).
\end{equation}
Therefore the value of the antenna pattern depends on the specific $k_{T}$
that we choose for the soft gluon, but in such a way that
$k_{T}^{2}\mathcal{R}$ tends to a finite limit as $k_{T} \to 0$.
Figure~\ref{fig:qqh_anrat_y2_kt1} illustrates this by taking the ratio of the
numerical $qq^{\prime}\rightarrow qq^{\prime}H$ antenna pattern with the
analytic $qq^{\prime}\rightarrow qq^{\prime}$ antenna pattern for $k_{T\,g} =
1$~GeV. The ratio is close to unity, except when the gluon rapidity is very
large. In this region the `soft' gluon carries a significant amount of energy
and begins to distort the overall kinematics. For numerical purposes only, as
a formal check that this effect is under control, we can set $k_{T\,g}$ to be
sufficiently (and artificially) small to make sure the analytic result is
recovered everywhere. Thus Fig.~\ref{fig:qqh_anrat_y2} shows the same ratio
for $k_{T\,g} =10^{-5}$~GeV -- no deviation from unity is now discernible.
Note that we will always use $k_{T\,g} = 1$~GeV in making predictions for the
antenna patterns using the numerical treatment. Since our ultimate aim is to
compare {\it two} numerically generated antenna patterns in signal to
background studies, the discrepancies at large gluon rapidity visible in
Fig.~\ref{fig:qqh_anrat_y2_kt1} will exactly cancel in the comparison.

\begin{figure}[htbp]
  \centering
  \epsfig{file=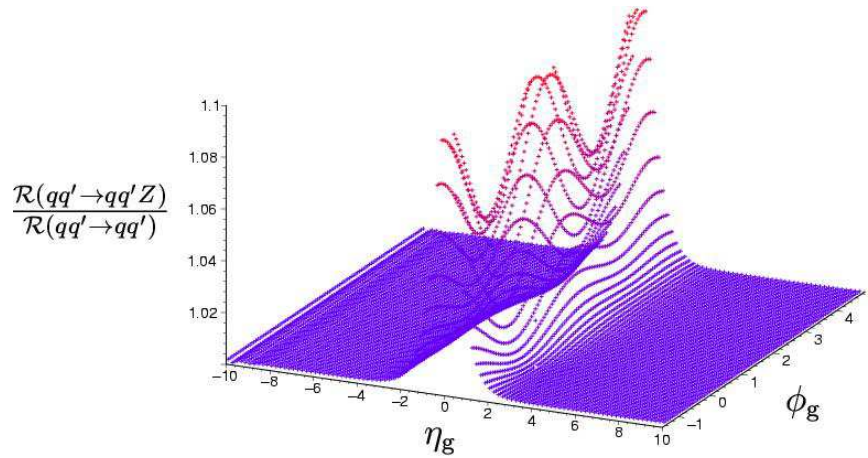,height=60mm}
  \caption{Ratio of numerical EW $qq^{\prime}\rightarrow qq^{\prime}Z$ to analytic $qq^{\prime}\rightarrow qq^{\prime}$ antenna patterns with $|\eta_{\textrm{jet}}|=2$ and $k_{T\textrm{g}}=10^{-5}$~GeV.}
  \label{fig:qqz_anrat_y2}
\end{figure}

\begin{figure}[htbp]
  \centering
  \epsfig{file=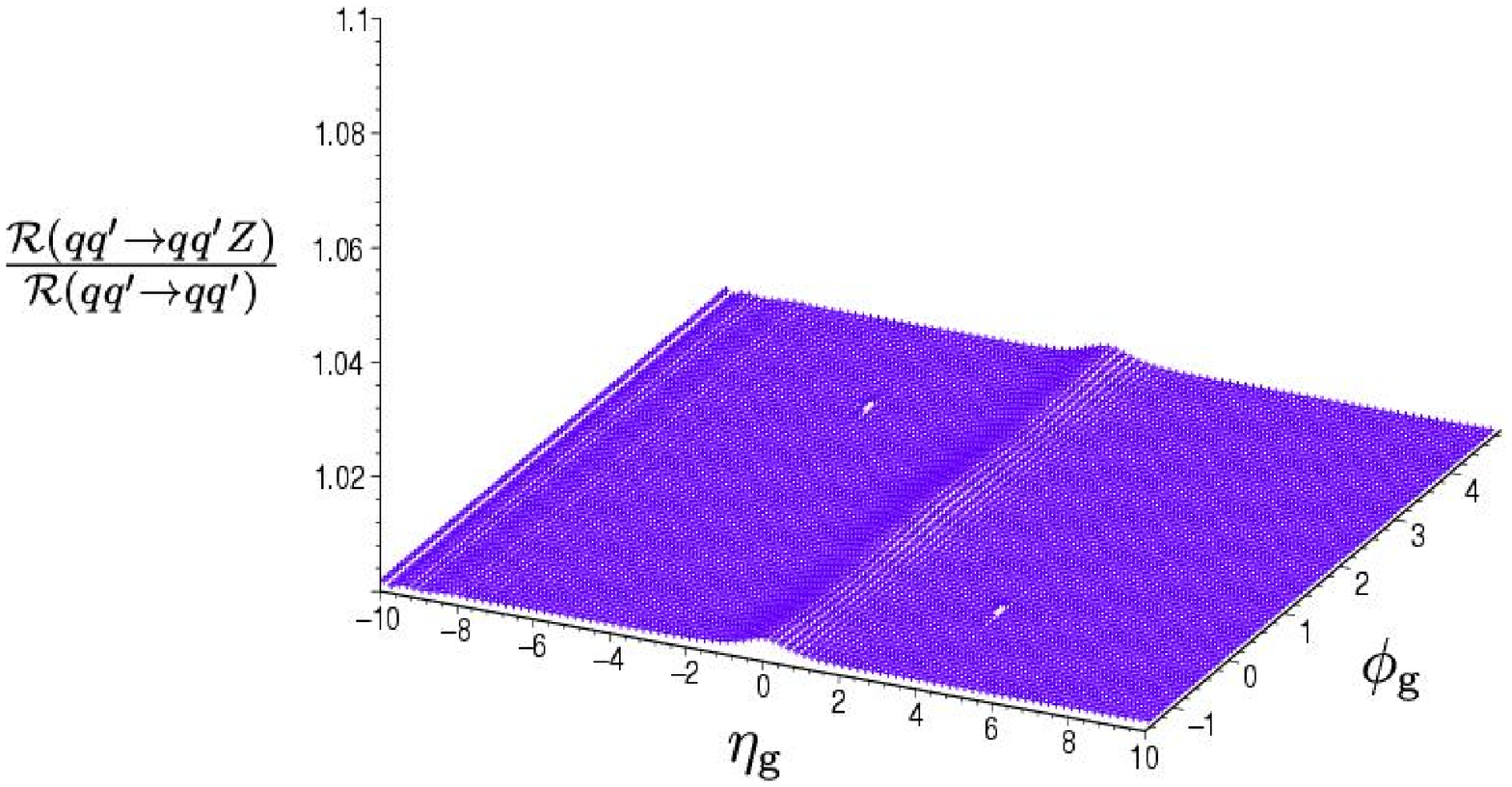,height=60mm}
  \caption{Ratio of numerical EW $qq^{\prime}\rightarrow qq^{\prime}Z$ to analytic $qq^{\prime}\rightarrow qq^{\prime}$ antenna patterns with $|\eta_{\textrm{jet}}|=4$ and $k_{T\textrm{g}}=10^{-5}$~GeV.}
  \label{fig:qqz_anrat_y4}
\end{figure}

As already pointed out, the antenna pattern for the full electroweak
$qq^{\prime}\rightarrow qq^{\prime}Z$ process is {\it not} given by the
simple analytic approximation, except when the jets are far forward. We can
now illustrate this using the numerical method. Thus
Figs.~\ref{fig:qqz_anrat_y2} and \ref{fig:qqz_anrat_y4} show the ratio of the
numerical electroweak $qq^{\prime}\rightarrow qq^{\prime}Z$ antenna pattern
with the analytic electroweak $qq^{\prime}\rightarrow qq^{\prime}$ antenna
pattern for the choice of $|\eta_{\textrm{jet}}|=2$ and
$|\eta_{\textrm{jet}}|=4$ with $k_{T}=10^{-5}$~GeV.  In the former case, the
agreement with the analytic antenna pattern is only at the 10\% level. The
discrepancy is due to the contribution of the $Z$-sstrahlung diagrams
(Fig.~\ref{fig:zew}b) in the numerical case. However, as one forces the quark
jets to be more forward the discrepancy decreases. Therefore, as long as we
require the jets to be forward (i.e. $|t|\ll \sqrt{s}$), the analytic
approximation is valid. \par

\begin{figure}[htbp]
  \centering
  \epsfig{file=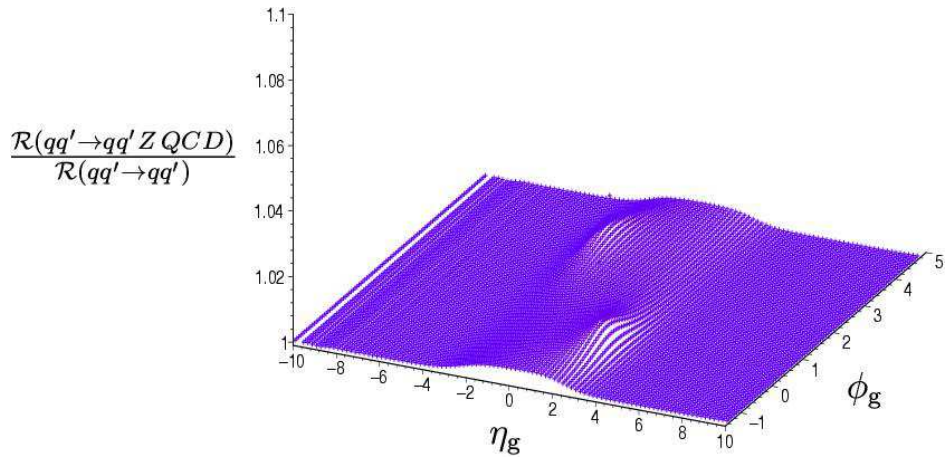,height=60mm}
  \caption{Ratio of numerical QCD $qq^{\prime}\rightarrow qq^{\prime}Z$ to analytic $qq^{\prime}\rightarrow qq^{\prime}$ antenna patterns with $|\eta_{\textrm{jet}}|=2$ and $k_{T\textrm{g}}=10^{-5}$~GeV.}
  \label{fig:qqzqcd_anrat_y2}
\end{figure}
\begin{figure}[htbp]
  \centering
  \epsfig{file=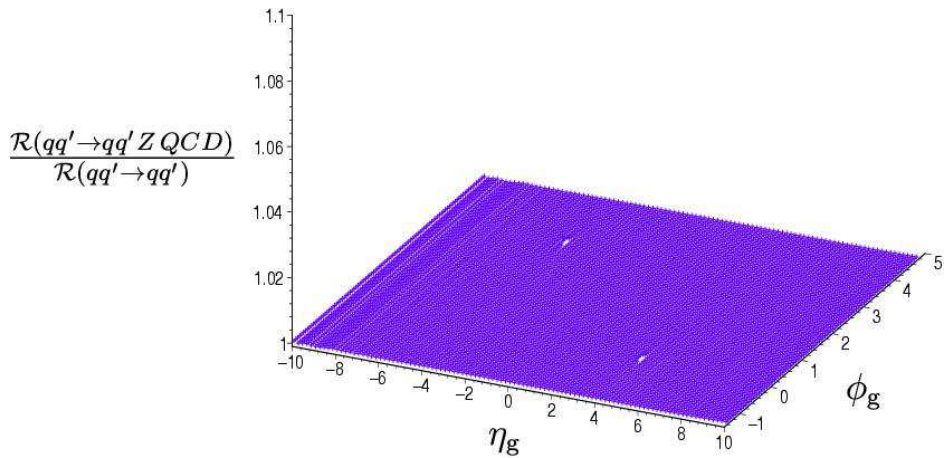,height=60mm}
  \caption{Ratio of numerical QCD $qq^{\prime}\rightarrow qq^{\prime}Z$ to analytic $qq^{\prime}\rightarrow qq^{\prime}$ antenna patterns with $|\eta_{\textrm{jet}}|=4$ and $k_{T\textrm{g}}=10^{-5}$~GeV.}
  \label{fig:qqzqcd_anrat_y4}
\end{figure}

Figures~\ref{fig:qqzqcd_anrat_y2} and \ref{fig:qqzqcd_anrat_y4} show the same
qualitative effect in the QCD mediated $Z$ production case. The deviation
from our approximation that $|t|\ll \sqrt{s}$ is noticeably less than in the
electroweak case. The reason for this is that in the electroweak case we are
kinematically disturbing a delicate interplay between the numerator and the
denominator in the term describing the colour connection between $p_{1}$ and
$p_{3}$
\begin{equation}
  \label{eq:13}
  [13]=\frac{p_{1}\cdot p_{3}}{(p_{1}\cdot k)(p_{3}\cdot k)}
\end{equation}
In particular, due to the smallness of the numerator, this contribution is
strongly suppressed for the radiation outside the narrow cones around the
directions of the incoming and outgoing partons. Contrast this with the QCD
$Z$ production case where the dominant colour connection is between $p_{1}$
and $p_{4}$
\begin{equation}
  \label{eq:14}
  [14]=\frac{p_{1}\cdot p_{4}}{(p_{1}\cdot k)(p_{4}\cdot k)}
\end{equation}
Here the numerator is not small. This cancellation is therefore more stable
and our kinematic disturbance has less effect.

\subsection{Numerical Antenna Patterns for Background Processes}
\begin{figure}[htbp]
  \centering
  \epsfig{file=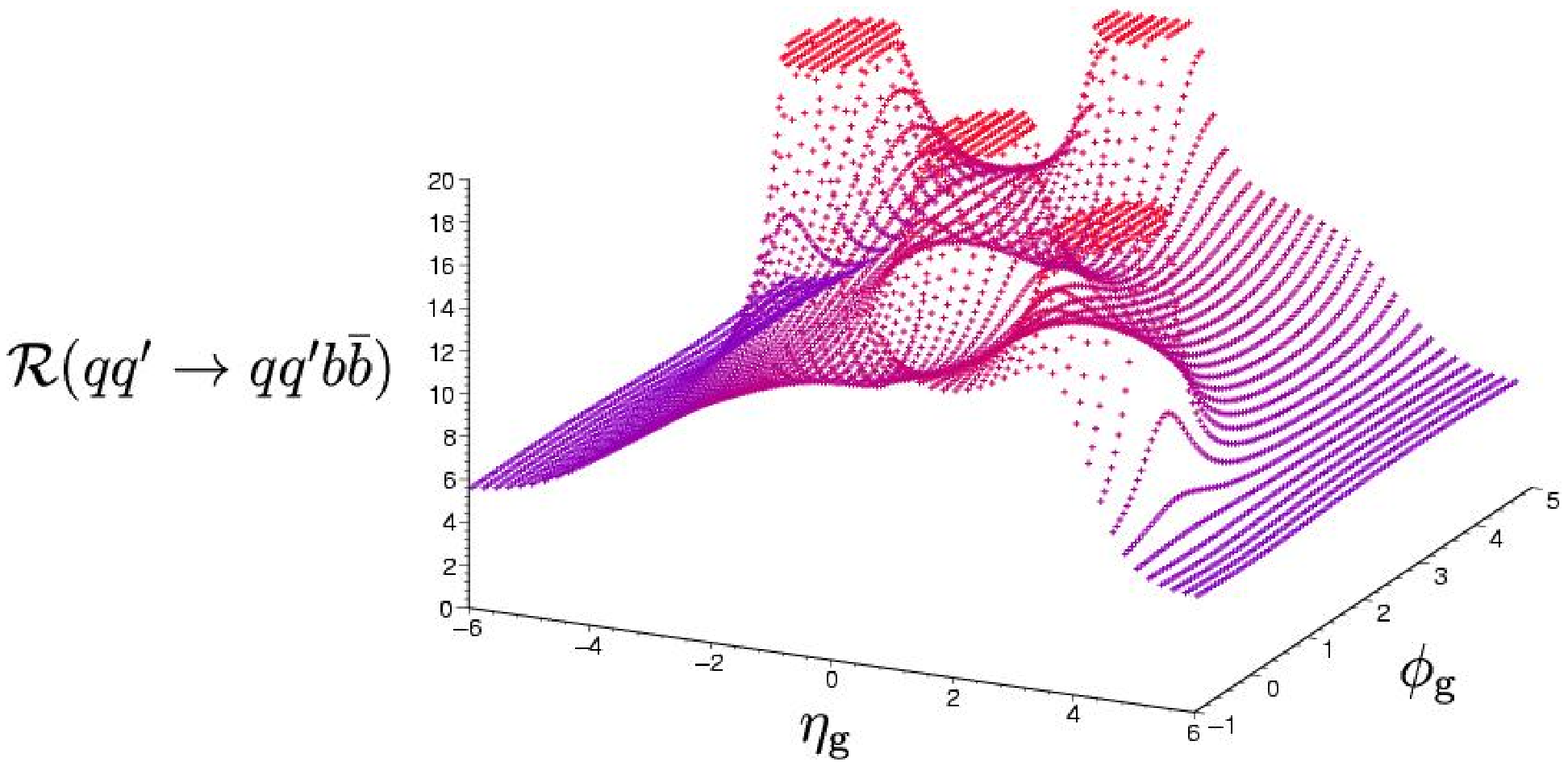,height=60mm}
  \caption{Numerical antenna pattern for $qq^\prime\rightarrow qq^\prime
  b\bar{b}$ with $|\eta_{\textrm{jet}}|=4$ and $k_{T\textrm{g}}=1$~GeV.}
  \label{fig:qq_qqbbb_y4}
\end{figure}

Figure~\ref{fig:qq_qqbbb_y4} shows the numerical antenna pattern for the QCD
mediated process $qq^\prime \rightarrow qq^\prime b\bar{b}$.
We will again focus mainly on the background process with initial state
quarks, to allow comparison with the signal processes. In any case, the
typical $\sqrt{\hat{s}}$ of the parton--level process is typically several
TeV at the LHC\footnote{For example, from Eq.~(\ref{eq:4mom}),
  $\sqrt{\hat{s}}\simeq m_{H}+2E_{T}\cosh{\eta_{\textrm{jet}}}\simeq 2.8$~TeV
  for $E_{T}=50$~GeV $m_{H}=120$~GeV and $\eta_{\textrm{jet}}=4$.}, so we are
working at high $x$ and quark initiated processes will dominate.  Therefore
the antenna patterns for the signal and background processes become identical
near the beam and final state $b$--quark directions, being dominated by the
(universal) collinear singularity for emission off quark lines.\par
\begin{figure}[htbp]
  \centering
  \epsfig{file=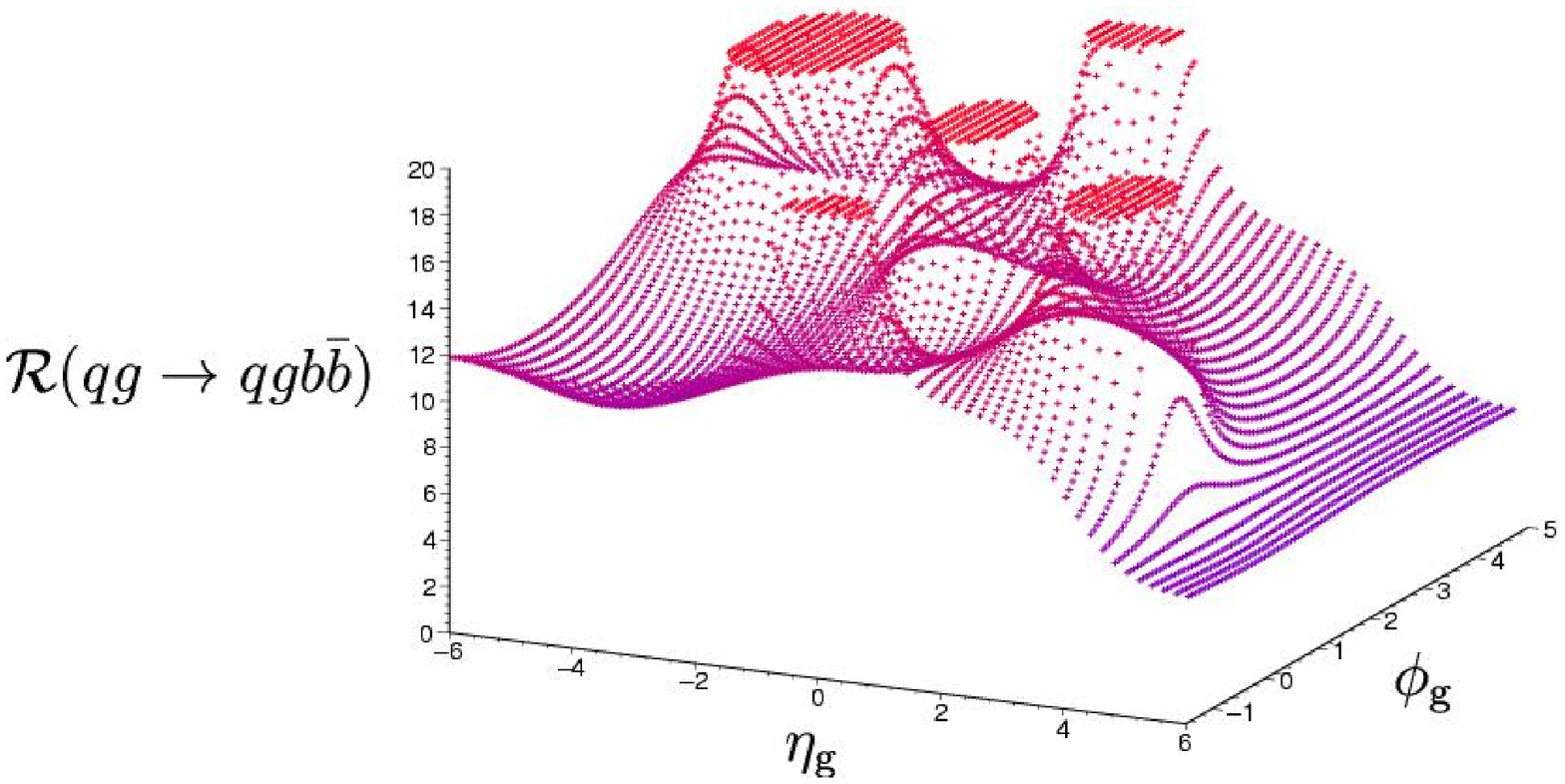,height=60mm}
  \caption{Numerical antenna pattern for $qg\rightarrow qgb\bar{b}$ 
    with $|\eta_{\textrm{jet}}|=4$ and $k_{T\textrm{g}}=1$~GeV.}
  \label{fig:qg_qgbbb_y4}
\end{figure}

Figure~\ref{fig:qg_qgbbb_y4} shows the radiation pattern for the background
QCD process $qg\rightarrow qgb\bar{b}$ with $|\eta_{\textrm{jet}}|=4$ and
$k_{T\textrm{g}}=1$~GeV. As expected, the pattern is much more complicated
than that for the signal $H$ or $Z$ production processes.  Colour strings can
now connect many more pairs of initial and final state particles, and the
overall level of radiation is higher as a result. However in the directions
of the incoming and outgoing partons, the distribution of soft radiation is
the same as that for the signal processes. Thus, in particular, the
distribution approaches $4C_F$ for large positive $\eta_{\textrm{g}}$, cf.
Eq.~(\ref{eq:eta_lim}).\par
\begin{figure}[htbp]
  \centering
  \epsfig{file=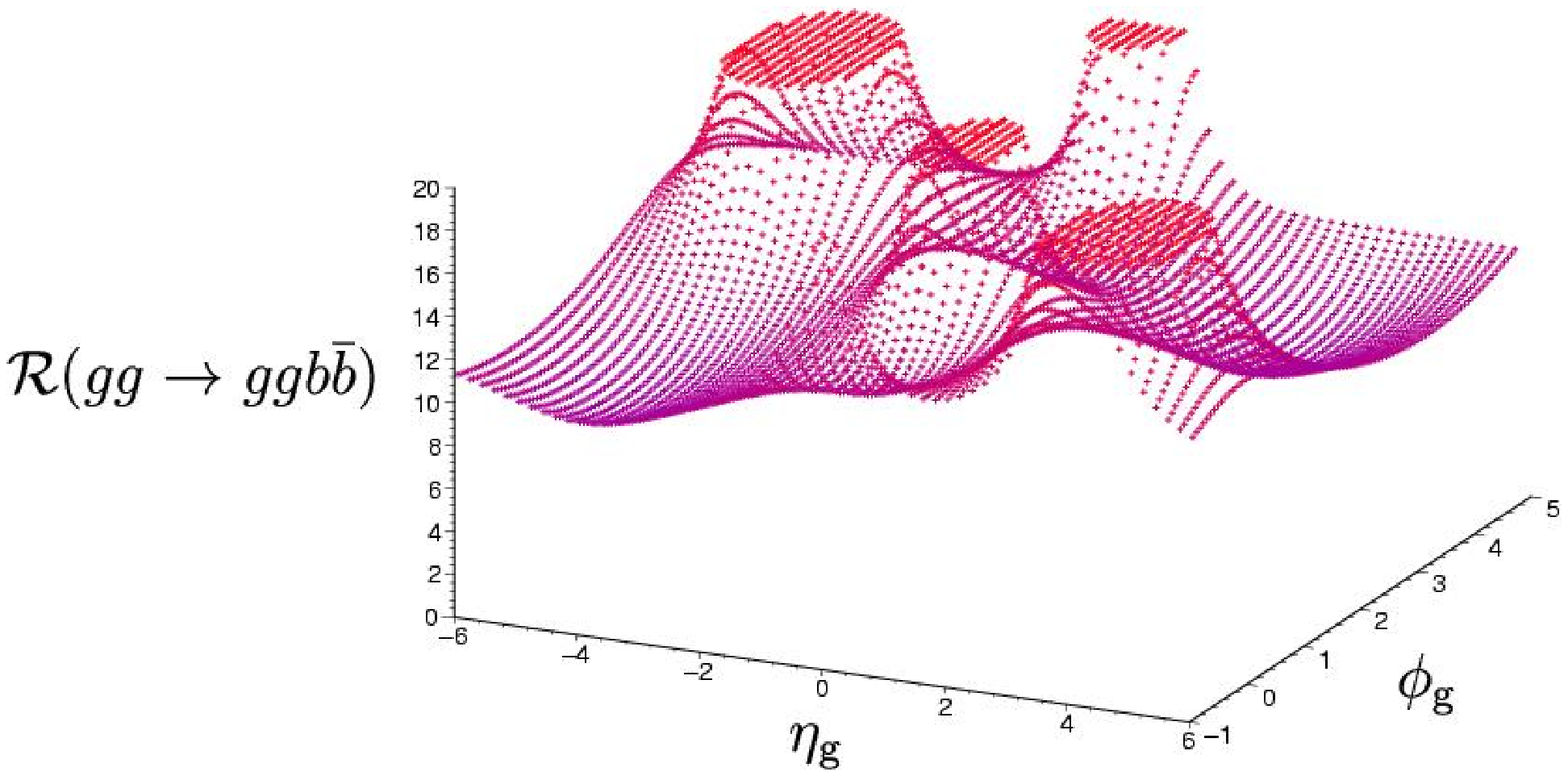,height=60mm}
  \caption{Numerical antenna pattern for $gg\rightarrow ggb\bar{b}$ 
    with $|\eta_{\textrm{jet}}|=4$ and $k_{T\textrm{g}}=1$~GeV.}
  \label{fig:gg_ggbbb_y4}
\end{figure}

\begin{figure}[htbp]
  \centering
  \epsfig{file=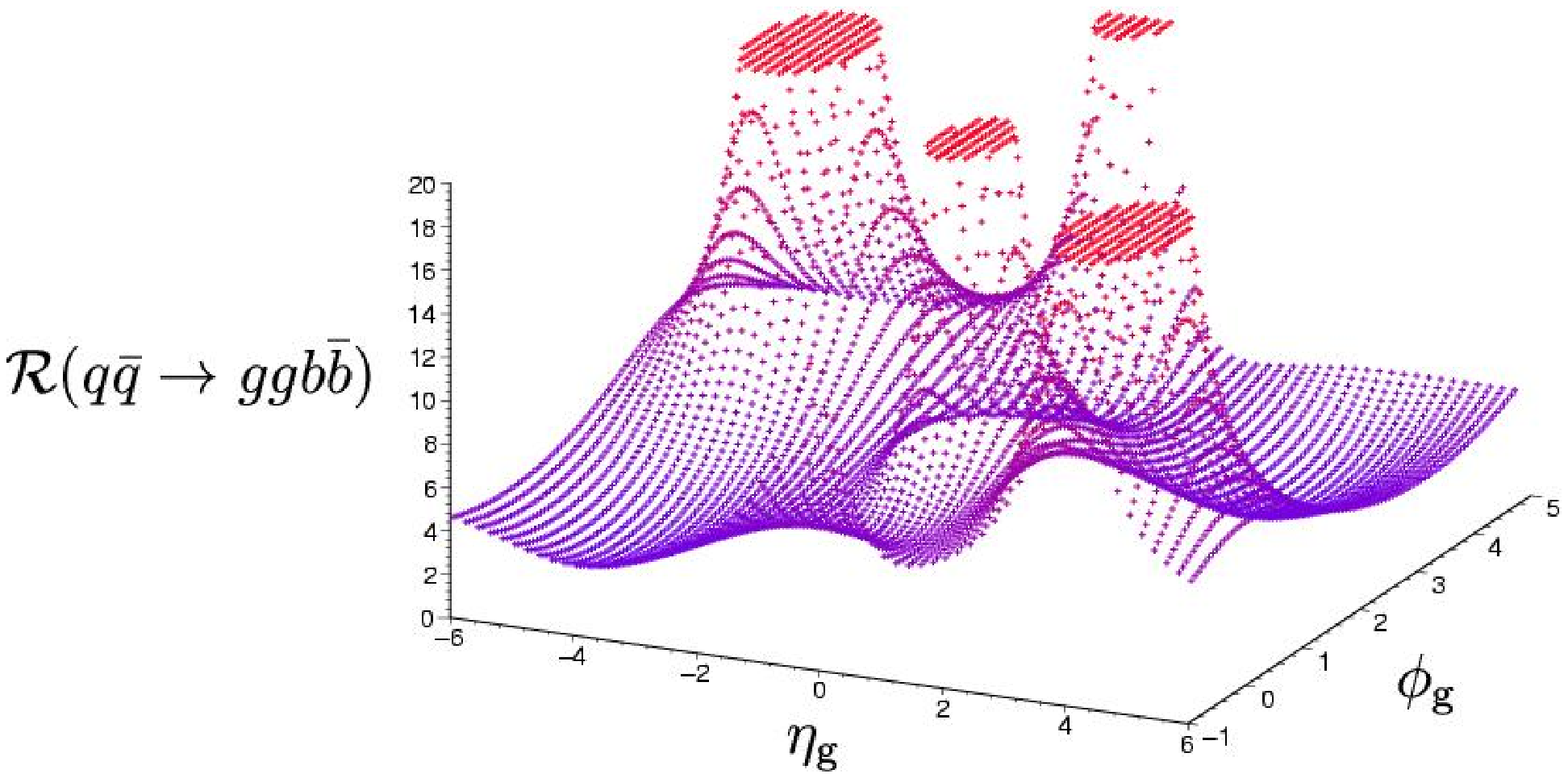,height=60mm}
  \caption{Numerical antenna pattern for $q\bar{q}\rightarrow ggb\bar{b}$ 
    with $|\eta_{\textrm{jet}}|=4$ and $k_{T\textrm{g}}=1$~GeV.}
  \label{fig:qqb_ggbbb_y4}
\end{figure}

\begin{figure}[htbp]
  \centering
  \epsfig{file=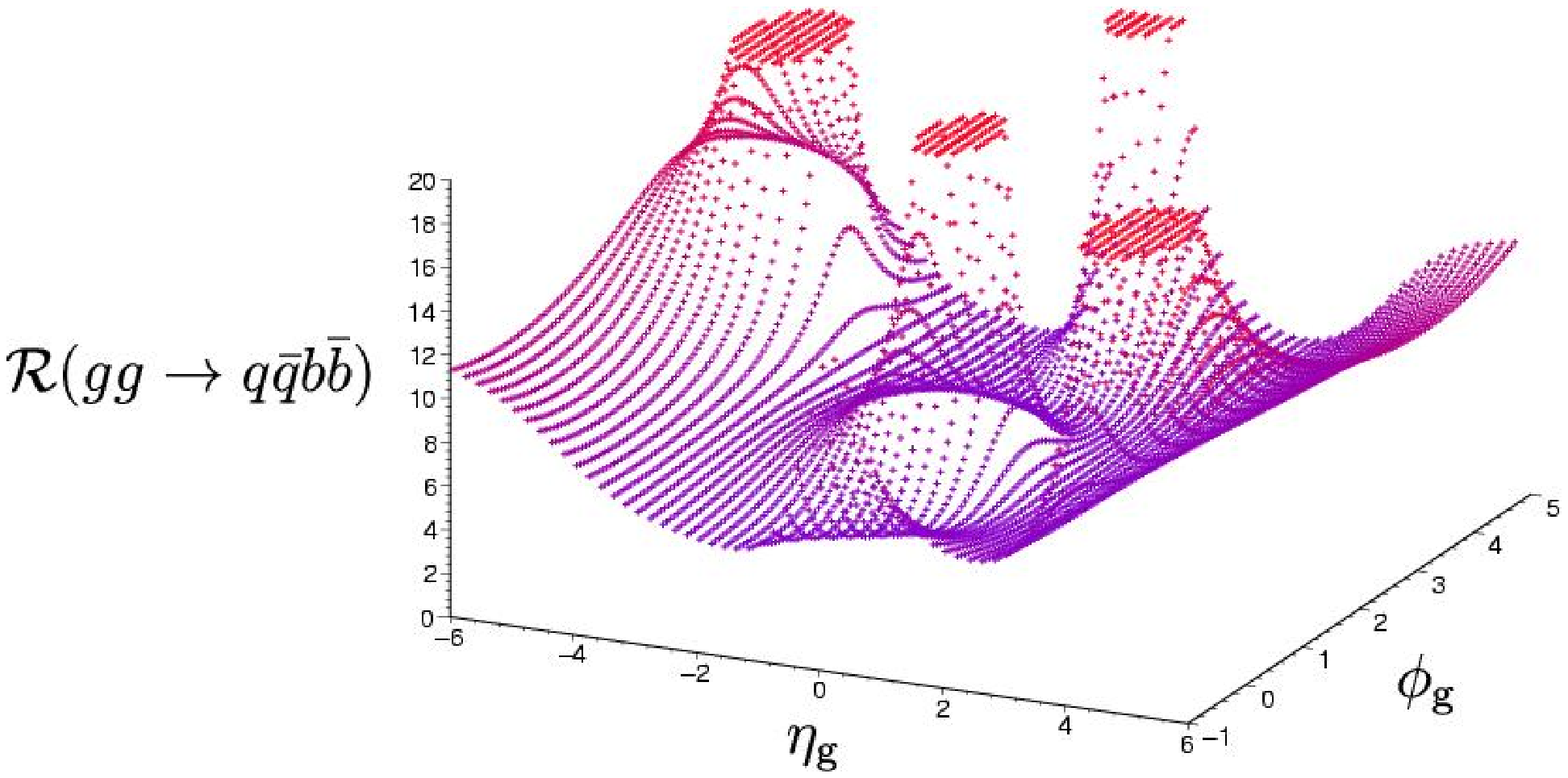,height=60mm}
  \caption{Numerical antenna pattern for $gg\rightarrow q\bar{q}b\bar{b}$ 
    with $|\eta_{\textrm{jet}}|=4$ and $k_{T\textrm{g}}=1$~GeV.}
  \label{fig:gg_qqbbbb_y4}
\end{figure}

 For completeness, we show in
Figs.~\ref{fig:qg_qgbbb_y4} -- \ref{fig:gg_qqbbbb_y4} the corresponding
antenna patterns for the other QCD $2 \to 2 + (b \bar b)$ processes.  The
most obvious differences are in the size of the distributions near the
incoming and outgoing partons, where the limiting $4C_F$ behaviour for
emission off quarks is replaced by $4C_A$ for emission off gluons.\par

\begin{figure}[htbp]
  \centering
  \epsfig{file=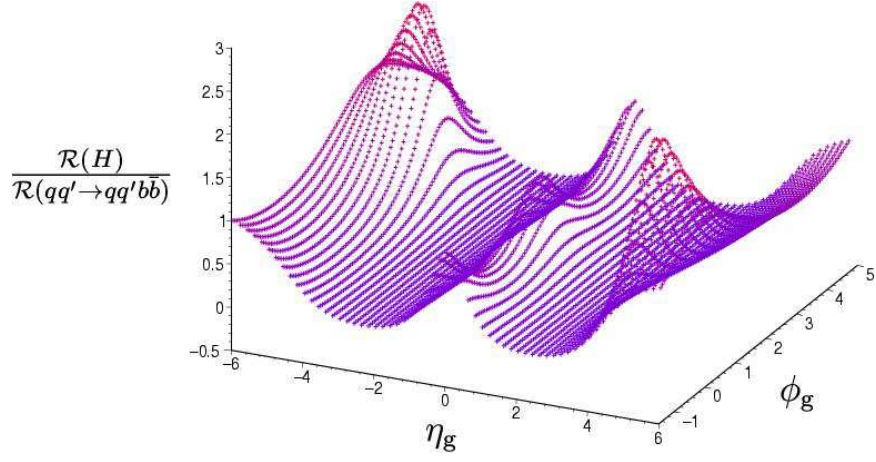,height=60mm}
  \caption{Ratio of numerical $qq^{\prime}\rightarrow
  qq^{\prime}H;\,H\rightarrow b\bar{b}$ to numerical $qq^{\prime}\rightarrow
  qq^{\prime}b\bar{b}$ with $|\eta_{\textrm{jet}}|=4$ and $k_{T\,g}=1$~GeV.}
  \label{fig:h_qcdbbb_num_k1gev}
\end{figure}

The interesting quantities are of course the {\it differences} between the
signals and backgrounds.  Figure~\ref{fig:h_qcdbbb_num_k1gev} shows the ratio
of numerical $qq^{\prime}\rightarrow qq^{\prime}H;\,H\rightarrow b\bar{b}$ to
numerical $qq^{\prime}\rightarrow qq^{\prime}b\bar{b}$ antenna patterns, for
the same typical kinematic configuration as before, i.e.
$|\eta_{\textrm{jet}}|=4$ and $k_{T\,g}=1$~GeV.  We see that the ratio (i)
falls to near zero between the central and forward particles (rapidity gap
effect), (ii) is larger than one between the final-state $b \bar b$ pair,
(iii) is larger than one between the forward jets and the beam (the [13] and
[24] connection in the signal), and (iv) approaches unity in the
forward/backward directions and at the locations (marked as arrows) of the
incoming and outgoing particles. Over the whole $(\eta,\phi)$ plot, the ratio
varies in size from a minimum of $0.03$ to a maximum of $2.3$, i.e. a factor
of $70$.  The corresponding ratio of the antenna patterns for the electroweak
$Z$ production and QCD background is of course very similar.\par
\begin{figure}[htbp]
  \centering
  \epsfig{file=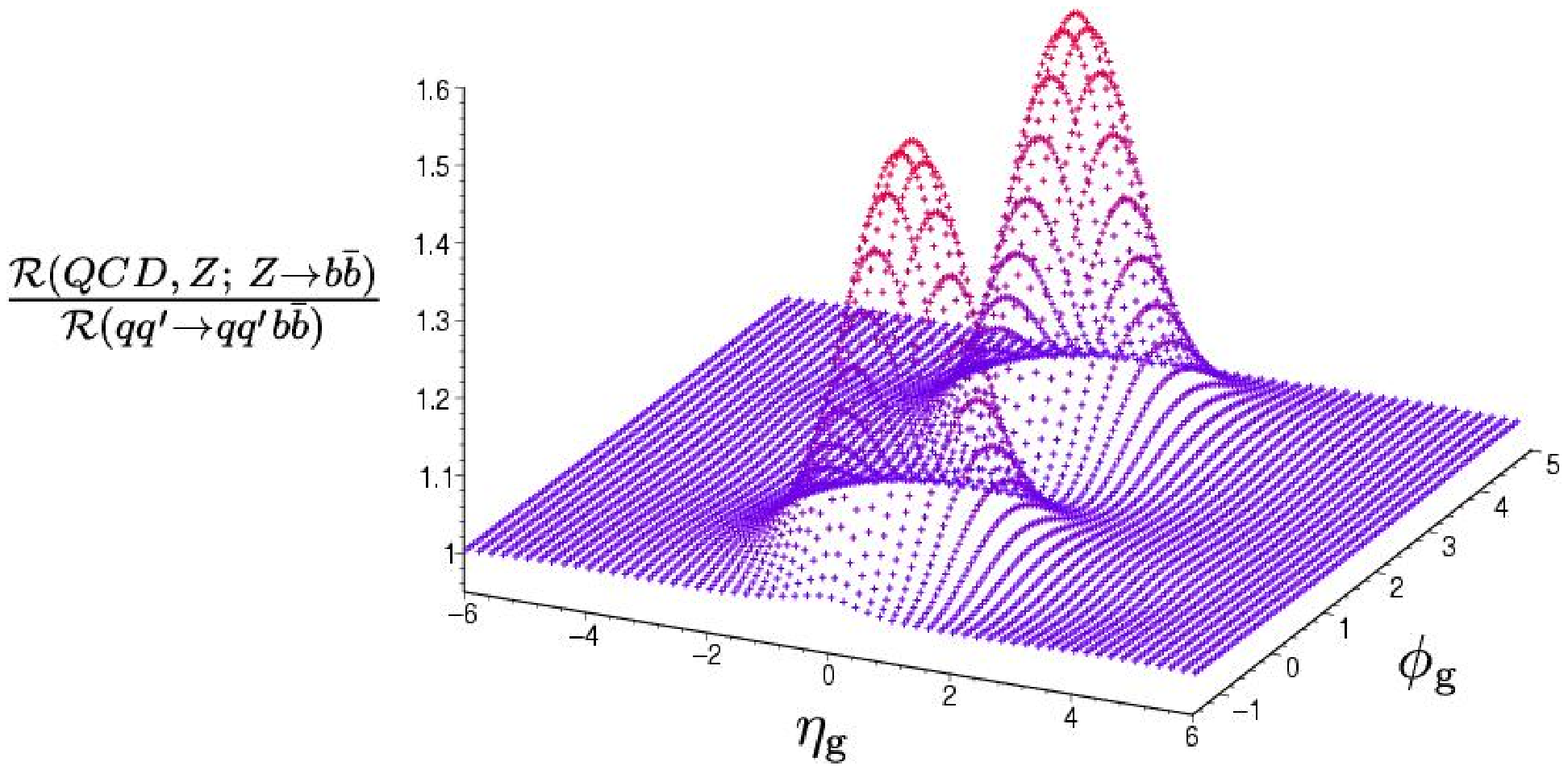,height=60mm}
  \caption{Ratio of numerical QCD $qq^{\prime}\rightarrow
    qq^{\prime}Z;\,Z\rightarrow b\bar{b}$ to numerical
    $qq^{\prime}\rightarrow qq^{\prime}b\bar{b}$ with
    $|\eta_{\textrm{jet}}|=4$ and $k_{T\,g}=1$~GeV.}
  \label{fig:qcdz_qcdbbb_num}
\end{figure}

 We next
consider (Fig.~\ref{fig:qcdz_qcdbbb_num}) the ratio of the QCD $Z$-production
and background $qq^{\prime}\rightarrow qq^{\prime}b\bar{b}$ antenna patterns.
There is much less structure here than there was in the corresponding Higgs
case -- note in particular that the rapidity gap dip between the forward and
central particles is absent\footnote{Note that by imposing the rapidity gap
  requirement to isolate the centrally produced system from the proton
  remnants, we would automatically cut off the colour connection between this
  system and the forward going partons. As shown in~\cite{Khoze:2002fa}, this
  allows us to substantially reduce the background contributions, though at
  the price of a reduction in the overall event rate (due to the notorious
  survival factors).}, and indeed that the ratio is close to one everywhere
except near the central $b$ jets.  In the $Z$ production case, there is
always a colour string connecting the $b$ and the $\bar b$, and this results
in the ratio increasing to a maximum of about 1.5 between these two
particles. This value has a weak dependence on the rapidities of the forward
jets. Figure~ shows the slice through Fig.~\ref{fig:qcdz_qcdbbb_num} at
$\eta_{\textrm{g}}=0$ as $|\eta_{\textrm{jet}}|$ is varied from 1 to 8.  The
ratio is always one at $ \phi_{\textrm{jet}} = \pi$ and $3\pi/2$, the
location of the $b$ and $\bar b$.


\begin{figure}[htbp]
  \centering
  \psfrag{x}[tl][bl]{$\phi_{\textrm{g}}|_{_{\eta=0}}$}
  \psfrag{y}[br][tl]{$\frac{{\cal R}(QCD,\,Z;\,\,Z\rightarrow b\bar{b})}{{\cal R}(qq^{\prime}\rightarrow
      qq^{\prime}b\bar{b})}$}
  \psfrag{Legend}[mr][bl]{Legend}
  \psfrag{8}[mm][mm]{$\eta=8$}
  \psfrag{n}[mm][mm]{$\eta=1$}
  \psfrag{z}[t][bl]{}
  \epsfig{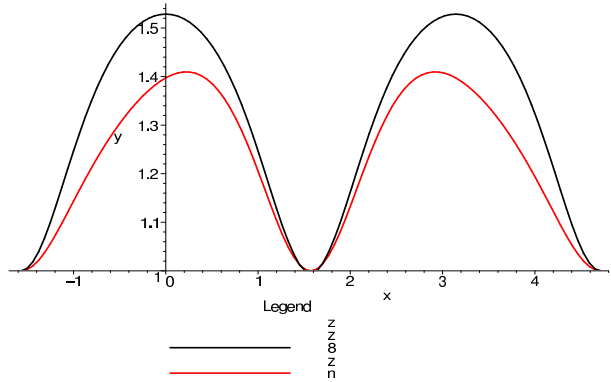}
  \caption{Slice in $\eta_{\textrm{g}}=0$ of Fig.~\ref{fig:qcdz_qcdbbb_num} 
    as the rapidity of the forward jets is varied.}
  \label{fig:sep_jet}    
\end{figure}
\section{Conclusions}
Hadronic radiation patterns can provide a useful additional tool enabling us
to improve the separation of Higgs production from the conventional
QCD-induced backgrounds. In this paper we have focused on the vector boson
fusion mechanism of Higgs production in the events with two forward tagging
jets. We find that the fairly simple analytical expressions reflecting
the coherent structure of QCD radiation off the multi-parton system (antenna
pattern) can serve quite successfully as a qualitative guide for the
more general numerical calculational technique, which in turn can be applied to
a large variety of complicated processes.\par The analysis presented here
should be regarded as a `first look' at the possibilities offered by hadronic
flow patterns in searching for the Higgs in vector boson fusion. Of course,
ultimately there is no substitute for a detailed Monte Carlo study including
detector effects. However the results presented here indicate that the
effects can be potentially large, and therefore that more detailed studies
are definitely worthwhile.


\def\Journal#1#2#3#4{{#1} {\bf #2}, #3 (#4)}

\def\NCA{Nuovo Cimento}
\def\NIM{Nucl. Instrum. Methods}
\def\NIMA{Nucl. Instrum. Methods {\bf A}}
\def\NPB{Nucl. Phys. {\bf B}}
\def\PLB{Phys. Lett. {\bf B}}
\def\PRL{Phys. Rev. Lett. }
\def\PRD{Phys. Rev. {\bf D}}
\def\ZPC{Z. Phys. {\bf C}}
\def\EPJC{Eur. Phys. J. {\bf C}}

\end{document}